\documentclass[aps,prd,twocolumn,superscriptaddress,floatfix,nofootinbib,nobibnotes]{revtex4-2}

\usepackage{epsfig}
\usepackage{bm}
\usepackage{amssymb}
\usepackage{amsmath}
\usepackage{color}
\usepackage{subfigure}
\usepackage{cancel}

\newcommand{\be}{\begin{equation}}
\newcommand{\ee}{\end{equation}}
\newcommand{\bea}{\begin{eqnarray}}
\newcommand{\eea}{\end{eqnarray}}

\usepackage{epsfig}
\usepackage{dcolumn}
\usepackage{bm}
\usepackage{amssymb}
\usepackage{amsmath}
\usepackage{hyperref}

\usepackage{graphicx}
\usepackage[dvipsnames]{xcolor}
\usepackage{color}
\definecolor{darkblue}{rgb}{0.0, 0.0, 0.55}
\definecolor{cite}{rgb}{0.0, 0.34, 0.25}
\definecolor{midgreen}{rgb}{0.52, 0.73, 0.4}

\usepackage{rotating}
\usepackage{bigints}
\usepackage{lineno}

\newcommand{\Lagr}{\mathcal{L}}

\begin{document}
\title{Bayesian location of the QCD critical point from a holographic perspective}
\date{\today}

\author{Mauricio~Hippert}
\email[Corresponding author: ]{hippert.mauricio@ce.uerj.br}
\affiliation{Illinois Center for Advanced Studies of the Universe \& Department of Physics,
University of Illinois at Urbana-Champaign, Urbana, IL 61801-3003, USA}
\affiliation{Instituto de Física, Universidade do Estado do Rio de Janeiro, Rua São Francisco Xavier, 524, Rio de Janeiro, RJ, 20550-013, Brazil}

\author{Joaquin~Grefa}
\affiliation{Department of Physics, University of Houston, Houston, TX 77204, USA}
\affiliation{Department of Physics, Kent State University, Kent, OH 44243, USA}

\author{T.~Andrew~Manning}
\affiliation{National Center for Supercomputing Applications, University of Illinois at Urbana-Champaign, Urbana, IL 61801, USA}

\author{Jorge~Noronha}
\affiliation{Illinois Center for Advanced Studies of the Universe \& Department of Physics,
University of Illinois at Urbana-Champaign, Urbana, IL 61801-3003, USA}

\author{Jacquelyn~Noronha-Hostler}
\affiliation{Illinois Center for Advanced Studies of the Universe \& Department of Physics,
University of Illinois at Urbana-Champaign, Urbana, IL 61801-3003, USA}

\author{Israel~Portillo~Vazquez}
\affiliation{Department of Physics, University of Houston, Houston, TX 77204, USA}

\author{Claudia~Ratti}
\affiliation{Department of Physics, University of Houston, Houston, TX 77204, USA}

\author{Romulo~Rougemont}
\affiliation{Instituto de F\'isica, Universidade Federal de Goi\'as,
Av. Esperan\c{c}a - Campus Samambaia,
CEP 74690-900, Goi\^ania, Goi\'as, Brazil}

\author{Michael~Trujillo}
\affiliation{Department of Physics, University of Houston, Houston, TX 77204, USA}

\begin{abstract}
A fundamental question in QCD is the existence of a phase transition at large doping of quarks over antiquarks. We present the first prediction of a QCD critical point (CP) from a Bayesian analysis constrained by first principle results at zero doping. We employ the gauge/gravity duality to map QCD onto a theory of dual black holes.  Predictions for the CP location in different realizations of the model overlap at one sigma.  
Even if many prior samples do not include a CP, one is found in nearly 100\% of posterior samples, indicating a strong preference for a CP.
\end{abstract}

\maketitle

\section{Introduction}

First principle lattice calculations have shown that quantum chromodynamics (QCD), the fundamental quantum field theory which accounts for $\sim 95\%$ of the visible matter in the Universe, undergoes an analytic crossover between phases  \cite{Aoki:2006we} when subjected to temperatures of $T\sim 10^{12}$ K and zero net baryon density. Such a smooth (though rapid) transition characterizes the change in the degrees of freedom of the theory from hadrons to a novel deconfined state of strongly interacting quarks and gluons. The conditions of temperature and density needed for this phenomenon occurred $\sim 20$ microseconds after the big bang \cite{Weinberg:2008zzc} and are constantly being reproduced, since the last decade, in ultrarelativistic heavy ion collisions at the Relativistic Heavy Ion Collider (RHIC) at Brookhaven National Laboratory, and the Large Hadron Collider (LHC) at CERN. These experiments have provided overwhelming evidence that quarks and gluons can behave collectively \cite{Heinz:2013th} as a new type of strongly interacting, nearly perfect fluid called the quark-gluon plasma (QGP) \cite{Gyulassy:2004zy,Shuryak:2004cy}. Since its first discovery in 2005, it quickly became clear that the QGP exhibits many unexpected features, being the smallest, hottest, and most perfect fluid ever observed.

The phase diagram of strongly interacting matter is still vastly unexplored: a quantitative description in a baryon-dense regime defies \textit{ab initio} lattice calculations due to the fermion sign problem, a fundamental obstacle of exponential complexity \cite{Troyer:2004ge} that affects any path integral representation of finite-density fermionic systems. Nevertheless, it is widely expected that increasing the imbalance between matter and anti-matter in this hot system will turn the crossover into a first-order phase transition, which would imply the existence of a critical point \cite{Stephanov:1998dy} in the QCD phase diagram. Understanding the emergence of critical phenomena in the theory of strong interactions is a fundamental challenge for both theory and experiment. Unlike condensed matter physics where doped systems can be investigated in equilibrium, heavy-ion experiments produce billions of collisions wherein highly dynamical quantum systems are formed that probe slightly different trajectories across the phase diagram. A scan of the phase diagram can be achieved by systematically decreasing the energy of the colliding ion beams: the second Beam Energy Scan (BESII) took place at RHIC during 2019-2021. New fixed target experiments will start operating in the next decade, allowing one to reach even larger densities. The QCD phase diagram at large densities is also crucial for the physics of neutron stars and neutron star mergers  \cite{Dexheimer:2020zzs,Almaalol:2022xwv,Lovato:2022vgq,Sorensen:2023zkk,MUSES:2023hyz}.

In the absence of a general mathematical framework, alternative approaches have been used to investigate the properties of dense fermionic systems. Our analysis makes use of the holographic gauge-gravity correspondence \cite{Maldacena:1997re,Gubser:1998bc,Witten:1998qj,Witten:1998zw}, a duality between a classical gravity theory in a 5-dimensional asymptotically anti--de Sitter (AdS$_5$) spacetime, and a strongly coupled quantum field theory which lives on its conformally flat 4-dimensional boundary. This approach has already been applied to quark-gluon plasma physics \cite{Son:2007vk,Gubser:2009md}: its main success is the natural emergence of nearly perfect fluidity \cite{Kovtun:2004de}, one of the most striking features of the QGP, in the strong coupling limit. 

In the holographic approach followed here, conformal invariance is broken by a real scalar field, which can be roughly understood as the running coupling of QCD; an additional U(1) gauge field is introduced to generate a baryonic charge and its corresponding chemical potential $\mu_B$ \cite{DeWolfe:2010he}. We find numerical solutions of the theory corresponding to thousands of charged black holes, each one of them dual to a point in the $T-\mu_B$ phase diagram of QCD. In previous applications \cite{Critelli:2017oub,Rougemont:2018ivt,Grefa:2021qvt,Grefa:2022sav}, we chose a specific functional form for the scalar field potential $V(\phi)$ and for the coupling between scalar and gauge fields $f(\phi)$, and we fixed their parameters to reproduce two crucial QCD quantities obtained through lattice simulations at $\mu_B=0$ for a system of 2+1 quark flavors: the equation of state \cite{Borsanyi:2013bia}, and the second-order fluctuation of the baryon charge \cite{Borsanyi:2021sxv}, which measures the response of the baryonic density to an infinitesimal change in the chemical potential. This led to a holographic prediction for the location of the QCD critical point \cite{Critelli:2017oub}, and to a holographic equation of state at finite $\mu_B$ \cite{Grefa:2021qvt} in quantitative agreement with state-of-the-art lattice results \cite{Borsanyi:2021sxv}. See \cite{Rougemont:2023gfz} for a comprehensive review.

In this manuscript, we employ the same classical gravity approach, but we choose two different functional forms for $V(\phi)$ and $f(\phi)$. We then use Bayesian inference to find the parameters that, for each functional form, yield the best description of the lattice QCD results. The prior distributions for $V(\phi)$ and $f(\phi)$ give rise to critical points scattered all over the phase diagram, or in some cases to no critical point at all. However, the posterior distributions that fall within the lattice QCD error bars, all yield holographic critical points that sit very close together, with 95\% confidence levels in the range $T = 101 - 108$ MeV and $\mu_B = 560 - 625$ MeV. %
Remarkably, this region falls within the extrapolated lattice QCD transition band between the hadron and quark-gluon plasma phases. Even more remarkably, the different functional forms predict compatible locations for the critical point, which are driven by the features of the lattice QCD results. 
We present a prediction of the collision energy needed to measure it in experiments.\\

\section{Holographic model}

The original formulation of the holographic gauge-gravity duality relates pure gravity in an asymptotically AdS spacetime in 5 dimensions to a strongly-coupled conformal super-symmetric Yang-Mills theory. Since QCD is not conformal, here we follow \cite{Gubser:2008ny,Gubser:2008yx,DeWolfe:2010he,DeWolfe:2011ts} and introduce a scalar dilaton field $\phi$ in the bulk theory to break conformal invariance. While there are three conserved charges in QCD, baryon number $B$, electric charge $Q$, and strangeness $S$, here we focus on the baryon number. Namely, we take a slice of the four-dimensional phase diagram corresponding to $\mu_S=0$ and $\mu_Q=0$.
A dual Abelian gauge field $A^\mu$ promotes the global $U(1)_B$ symmetry associated with baryon-number conservation to a local symmetry in the bulk.

\subsection{Model construction}

  The action of the 5-dimensional gravitational theory for the so-called Einstein-Maxwell-dilaton (EMD) model is given by
\begin{eqnarray} \label{eq:action}
    S&=&\int_{\mathcal{M}_5} d^5 x \Lagr = \frac{1}{2\kappa_{5}^{2}}\int_{\mathcal{M}_5} d^{5}x\sqrt{-g}\times
    \nonumber\\
    &\times&\left[R-\frac{(\partial_\mu \phi)^2}{2}-V(\phi)-\frac{f(\phi)F_{\mu\nu}^{2}}{4}\right],
\end{eqnarray}
where, on the right-hand side, the different terms are the Einstein-Hilbert action, the dilaton field kinetic term, the dilaton potential, and the Maxwell action, with $F_{\mu\nu} \equiv \partial_\mu A_\nu - \partial_\nu A_\mu$. The $F_{\mu\nu}^2$ term is scaled by a function $f(\phi)$ of the dilaton field, which couples the renormalization group flow to the baryon current without breaking $U(1)_B$. In Eq.~\eqref{eq:action}, $\kappa_5^2=8\pi G_5$ is Newton's constant in five-dimensional spacetime.

{Considering homogeneous and isotropic black holes in equilibrium, and taking convenient coordinate and gauge choices, we pick the following forms for the EMD fields \cite{DeWolfe:2010he}
\begin{align}
    &ds^2 = e^{2A(r)}[-h(r)dt^2+d\vec{x}^2]+\frac{dr^{2}}{h(r)},&
    \label{eq:metric}\\
    & A^\mu(x)=\delta_0^\mu \Phi(r), \qquad \phi(x)=\phi(r),&
    \label{eq:gauge}
\end{align}
where $r$ is the holographic radial coordinate specifying the position along the extra dimension. 
A black hole horizon deep in the bulk is responsible for inducing a thermal behavior at the dual quantum field theory living at the boundary. The black hole horizon location $r=r_H$ is specified by the largest root of the blackening function, $h(r_H)=0$. 
}

\subsection{Holographic potentials}

For the functions $V(\phi)$ and $f(\phi)$ we do not have a systematic expansion to regulate their functional forms. Yet, in the past, we proposed an \textit{Ansatz} that allowed us to reproduce lattice QCD results and predict thermodynamic observables and transport coefficients \cite{Critelli:2017oub,Rougemont:2018ivt,Grefa:2021qvt,Grefa:2022sav}. Here, we test different functional forms and perform, for the first time in this field, a Bayesian analysis to fix their parameters to reproduce the lattice QCD results at $\mu_B=0$. Our goal is to investigate how these choices affect the location of the predicted critical point and how the lattice QCD features affect these predictions.

The first functional form we propose is a polynomial-hyperbolic \textit{Ansatz} (PHA), which has often been used for this kind of model in the past \cite{Gubser:2008ny,Gubser:2008yx,DeWolfe:2010he,DeWolfe:2011ts,Finazzo:2014cna,Rougemont:2015wca,Rougemont:2015ona,Finazzo:2015xwa,Rougemont:2017tlu,Critelli:2017oub,Rougemont:2018ivt,Grefa:2021qvt,Grefa:2022sav,Cai:2022omk,Li:2023mpv,Jokela:2024xgz}. We propose the following functions:
   \begin{equation}\label{eq:hyperpoly_V}
 V(\phi) = -12\cosh(\gamma\,\phi)+b_2\,\phi^{2}+b_4\,\phi^{4}+b_6\,\phi^{6},
\end{equation}
\begin{multline}
\label{eq:hyperpoly_f}
 f(\phi) = \frac{\mathrm{sech}(c_{1}\phi+c_{2}\phi^{2}+c_{3}\phi^{3})}{1+d_{1}}+\frac{d_{1}}{1+d_{1}}\mathrm{sech}(d_{2}\phi).
\end{multline}
These functional forms are similar to the ones proposed by some of us in Ref.~\cite{Critelli:2017oub}, but they also include the cubic term in the argument of the hyperbolic secant, which was introduced in Ref.~\cite{Cai:2022omk}. The last term in Eq.~\eqref{eq:hyperpoly_f} replaces the pure exponential of Ref.~\cite{Cai:2022omk}. We note that in Ref.~\cite{Cai:2022omk}, $b_4$, $c_1$ and $c_2$ were not considered.

The functional forms in Eqs.\ \eqref{eq:hyperpoly_V} and \eqref{eq:hyperpoly_f} exhibit distinct features, such as exponential slopes and plateaus, at different values of $\phi$. However, these features are not uniquely driven by a specific coefficient in the above functions. For this reason, we propose a new parametric \textit{Ansatz} (PA) for them \begin{equation}\label{eq:parametric_V}
 V(\phi) = -12\cosh\left[\left(\frac{\gamma_1\,\Delta\phi_V^2 + \gamma_2 \,\phi^2}{\Delta \phi_V^2 + \phi^2}\right) \phi\right],
\end{equation}
\begin{multline}\label{eq:parametric_f}
 f(\phi) = 1 - (1-A_1) \left[\frac{1}{2} + \frac{1}{2}\tanh\left(\frac{\phi - \phi_1}{\delta \phi_1}\right)\right] \\ 
 - A_1\left[\frac{1}{2} + \frac{1}{2}\tanh\left(\frac{\phi - \phi_2}{\delta \phi_2}\right)\right].
\end{multline}
This \textit{Ansatz} has the advantage of having parameters that are easier to interpret since they now control the features described above. Besides being able to produce an EMD model that mimics lattice thermodynamics at zero chemical potential, this \textit{Ansatz} will provide further information regarding the dependence of the predicted critical point on the choice for $V(\phi)$ and $f(\phi)$. We note that the form for $f(\phi)$ is similar to the one proposed in Ref.~\cite{Knaute:2017opk}.\\

Both \textit{Ans\"atze} are such that $V(0)=-12$, so that at the $r\to \infty$ boundary, the Ricci scalar approaches $-20$, in units of an energy scale $\Lambda$ squared (this is required for having asymptotically AdS$_5$ background solutions, in consonance with the holographic dictionary). This scale is used to convert model results to physical units and is a free parameter of our model \cite{Critelli:2017oub}.

\subsection{Numerical solutions}

Using Eqs.~\eqref{eq:metric}, \eqref{eq:gauge} and extremizing the action in Eq.~\eqref{eq:action} yields a set of coupled second-order ordinary differential equations, which, in general, require numerical methods to be solved \cite{DeWolfe:2010he}. 
By conveniently changing coordinates before implementing a numerical solution, the horizon data to be chosen is reduced to the spectification of the values of the dilaton field and the radial derivative of the $U(1)$ gauge field at the horizon \cite{Critelli:2017oub}:
\begin{align}
 \phi_0 &\equiv \phi(r_{H})\,,
 &
 \Phi_1 &\equiv  \Phi'(r_{H})\,,
\end{align}
which provide initial conditions for the numerical solution of the equations of motion in the bulk. 
That is, each $(\phi_0,\,\Phi_1)$-pair specifies a given (equilibrium) thermodynamic state of the system at the boundary.  

After numerically solving the equations of motion, one can again change coordinates to a form such that standard holographic formulas can be applied to extract thermodynamic quantities in the boundary theory \cite{Critelli:2017oub}. 
This is done by analyzing the asymptotic behavior of the fields near the asymptotically AdS$_5$ boundary at $r\to \infty$. 

Here, we solve the equations of motion and extract thermodynamic quantities for different $(\phi_0,\,\Phi_1)$-pairs using software developed within the MUSES framework, which will be detailed elsewhere. 

\subsection{Critical point location}

\begin{figure}[t]
    \centering
    \includegraphics[width=\linewidth]{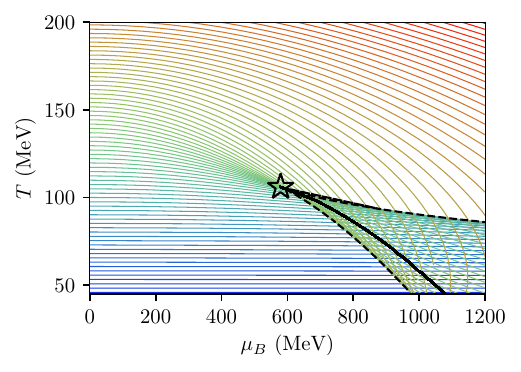}\\
    \caption{Example of critical point location by finding the point in the $(T,~\mu_B)$ phase diagram where the lines of constant $\phi_0$ at increasing $\Phi_1$ cross for the first time.}
    \label{Fig:CP}
\end{figure}

To understand what happens at the critical point, it is instructive to plot lines of constant $\phi_0$ (for varying $\Phi_1$) on the phase diagram. 
These lines are shown in Fig.\ \ref{Fig:CP}, where the color code indicates the value of $\phi_0$. The CP is marked by a star, the first order line is indicated as the solid black line, and the spinodal lines are represented as dashed black lines. 

We see that the constant $\phi0$ lines start off parallel at $\mu_B=0$, but as we increase $\mu_B$ their behavior changes, leading to crossings between them. 
These crossings happen because between the spinodal curves there are three possible thermodynamic states corresponding to a single $(T, \mu_B)$-pair: one stable, one metastable and one thermodynamically unstable. 
Since these states correspond to different values of $\phi_0$, up to three different lines can intersect in this region, which starts precisely at the critical point.

We thus construct a CP-locating algorithm to automatically find the intersection between these lines, and use it to locate the critical point for each prior and posterior curve for $V(\phi)$ and $f(\phi)$.

\begin{table*}[t]
    \centering
    \begin{tabular}{|c||c|c|}
    \hline
    \multicolumn{3}{|c|}{PHA \textit{Ansatz}}\\
    \hline\hline
         Parameter & min & max  \\
    \hline\hline
        $\Lambda$ & 800 MeV & 1400 MeV\\
    \hline
        $\kappa_5^2$ & 9.0 & 15.0\\
    \hline\hline
        $\gamma$ & 0.5682 & 0.6500\\
    \hline
        $b_2$ & -0.05 & 0.65\\
    \hline
        $b_4$ & -0.150 & -0.015\\
    \hline
        $b_6$ & -0.00200 & 0.00169\\
    \hline\hline
        $c_1$ & -0.035 & 0.100\\
    \hline
        $c_2$ & 0.1 & 1.5\\
    \hline
        $c_3$ & 0.0 & 1.0\\
    \hline
        $d_1$ & 0.0 & 2.5\\
    \hline
     \phantom{(J)}   $d_2$ {(J)} & 3 & 10000\\
    \hline
    \end{tabular}\hspace{2cm}%
    \begin{tabular}{|c||c|c|}
    \hline
    \multicolumn{3}{|c|}{PA \textit{Ansatz}}\\
    \hline\hline
         Parameter & min & max  \\
    \hline\hline
        $\Lambda$ & 400 MeV & 1400 MeV\\
    \hline
        $\kappa_5^2$ & 9.0 & 15.0 \\
    \hline\hline
       $\gamma_1$ & 0.40 & 0.57 \\
       \hline
       $\gamma_2$ & 0.50 &0.68 \\
    \hline
      $\Delta \phi_V$ & 1.5 & 3.0 \\
      \hline\hline
       $A$ & 0.25 & 0.50\\
    \hline
       $\phi_1$ & -0.1 & 0.5\\
    \hline
       \phantom{(J)} $\delta\phi_1$ {(J)} & $10^{-5}$ & 0.3\\
    \hline
       $\phi_2$ & 0.8 & 4.5 \\
    \hline
       $\delta\phi_2$ & 0.2 & 4.0\\
    \hline
    \end{tabular}    
    \caption{Prior ranges for parameters in the PHA (left) and PA (right) models. 
    The (J) marks parameters for which we have used \emph{Jeffreys} priors---i.e., prior distributions that are uniform over the logarithm of these parameters.}
    \label{tab:priors}
\end{table*}

\section{Bayesian analysis}

We wish to use lattice QCD constraints to draw probabilistic predictions for the QCD critical point, according to lattice QCD uncertainties, within the framework of our holographic EMD model. 
This is done by using Bayesian inference to define a posterior probability distribution over the parameter space of our model for each of the \textit{Ans\"atze} above. 
By sampling a large number of parameter sets from this posterior distribution and computing predictions for each of them, we are able to extract a probability distribution for the location of the QCD critical point.

To assign probabilities to sets of parameters given lattice QCD constraints, we use Bayes' theorem
\begin{equation}
\label{eq:bayes}
    P(\vec \theta| \vec d) = \frac{P(\vec d| \vec \theta)\times P(\vec \theta)}{P(\vec d)}\,,
\end{equation}
where $\vec d$ denotes the lattice QCD constraints and $\vec \theta$ denotes a parameter set, including not only the parameters in the holographic potentials $V(\phi)$ and $f(\phi)$, but also the gravitational constant $\kappa_5^2$ and the  conversion scale $\Lambda$. %
On the left-hand side of Eq.~\eqref{eq:bayes}, $P(\vec\theta|\vec d)$ is the desired probability distribution over parameter space after imposing the constraints $\vec d$, known as the posterior. 
On the right-hand side, $\mathcal{L}(\vec\theta)\equiv P(\vec d|\vec \theta)$, known as the likelihood, represents the probability of reproducing the constraints $\vec d$ given $\vec \theta$. 
The prior distribution $P(\vec \theta)$ is the probability over parameter space \emph{before} imposing the constraints $\vec d$, therefore representing prior assumptions and knowledge. 
Finally, $P(\vec d)$ is independent of $\vec \theta$ and can be seen as a normalization factor. 

To sample parameter sets from the posterior distribution $P(\vec \theta|\vec d)$ given by Eq.~\eqref{eq:bayes}, we employ Markov Chain Monte Carlo (MCMC) methods. 
That is, starting from a number of points in parameter space, we successively make random updates to them, with transition probabilities chosen such that, after a large number of iterations, they become distributed according to $P(\vec \theta|\vec d)$. 
In particular, we use differential evolution MCMC (DE-MCMC) \cite{2006S&C....16..239T}, which mitigates issues from strong correlations between different parameters. 
Details on our DE-MCMC implementation can be found in Appendix~\ref{app:demc}.

\begin{figure*}[t]
    \includegraphics[width=0.68\linewidth]{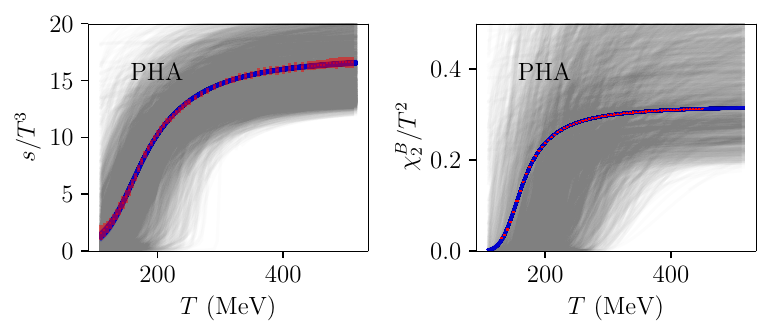}%
    \includegraphics[width=0.31\linewidth]{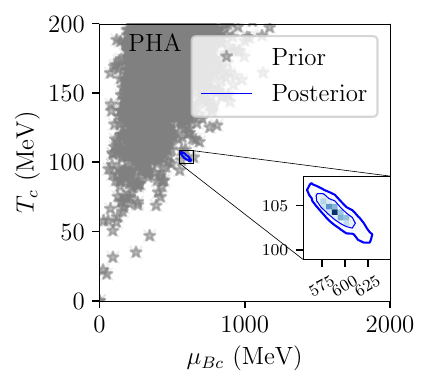}\\%
    \includegraphics[width=0.68\linewidth]{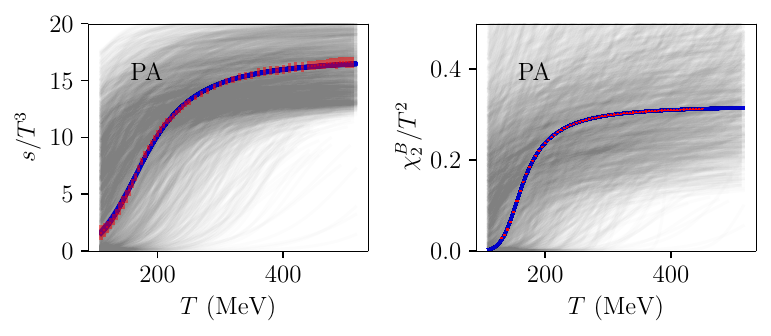}%
    \includegraphics[width=0.31\linewidth]{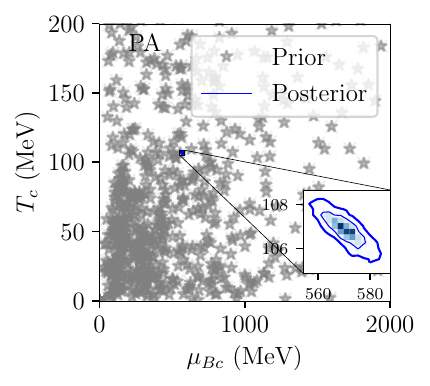}%
    \caption{
    Results for sample equations of state from the prior (gray) and posterior (blue) distributions, for the PHA (top) and PA (bottom) \textit{Ans\"atze}. 
    Left: entropy density, normalized by temperature cubed, as a function of temperature. 
    Center: second order baryon number susceptibility {normalized by temperature squared}, as a function of temperature. 
    Right: predictions for the QCD critical point for the prior (gray stars) and posterior samples (blue histogram and contours). 
    In the left and center panels, the red points
with error bars are the lattice results from Refs.\ \cite{Borsanyi:2013bia,Borsanyi:2021sxv}. 
    In the rightmost panels, the blue lines visible in the inset represent 68\% and 95\% confidence levels for the posterior distribution.   
}
    \label{fig:priors}
\end{figure*}

\subsection{Prior distribution}

For simplicity, we take uniform prior distributions over most parameters, 
within the ranges shown in Table~\ref{tab:priors}. 
The only exceptions are the parameters $d_2$, in the PHA model, and $\delta\phi_1$, in the PA model, for which we employ Jeffreys prior distributions (uniform over the logarithm of these parameters), also within the ranges indicated in Table~\ref{tab:priors}, where they are marked by a `(J)'\footnote{The Jeffreys prior is indicated in cases where there is uncertainty on the order of magnitude of a model parameter.}.

In choosing the parameter ranges in Table~\ref{tab:priors}, we have tried to be as unprejudiced as possible. 
For simplicity, we choose ranges for each parameter independently, forming a hyperrectangle in parameter space. 

As a starting point for the Markov Chain Monte Carlo, prior samples are initially drawn in a Latin hypercube configuration using library \texttt{pyDOE},
but we remove from the prior any points in parameter space which violate the condition  $-4\leq V''(0) <0$ \cite{Breitenlohner:1982bm,Breitenlohner:1982jf}, so that the dilaton field is both stable and relevant in the infrared. 
We sample with 500 points per parameter for the PHA model, and 200 points per parameter for the PA model\footnote{More points are initially sampled for the PHA model to compensate for the fact that some of the models in the prior ranges turn out to be unstable and are not evaluated.}.

The partially gray lines in the leftmost and middle panels of Fig.~\ref{fig:priors} display the entropy density and the second baryon susceptibility $\chi_2^B$ for samples of the prior 
utilized to start the DE-MCMC algorithm. 
The rightmost panels show the spatial distribution of critical points in the $(T,~\mu_B)$-plane corresponding to these samples of the priors. 
The top and bottom panels correspond to the PHA and PA models, respectively. 
It is evident that priors for the PA version of the EMD model cover a wider range for the equation of state, especially for  $\chi_2^B$.  While $\sim 20\%$ of the prior sample does not produce a critical point at all for the PA model,\footnote{About $30\%-50\%$ of the prior sample for the PA model lacks a critical point, but some of it is penalized in our analysis, for missing  points---i.e., not covering all the temperatures in the lattice results due to computational or model limitations---or for having a phase transition at $\mu_B=0$. If penalized realizations are removed, the proportion of the sample without a critical point is reduced to $\sim 20\%$.}
critical points found in this sample are 
scattered over a very wide region in the phase diagram. 
On the other hand, the prior for the PHA version of the model comparatively produces critical points that are concentrated in a smaller region of the phase diagram.

\subsection{Lattice QCD constraints}

In Eq.~\eqref{eq:bayes}, lattice QCD constraints are imposed via the likelihood function $\mathcal{L}(\vec\theta)\equiv P(\vec d|\vec \theta)$.  
The likelihood function $\mathcal{L}(\vec \theta)$ represents the probability of obtaining our choice of lattice QCD results from a given realization of our model, with parameters $\vec \theta$. 
This probability depends on the uncertainties, and on how the error on the lattice QCD results is distributed. 
We assume a Gaussian likelihood, corresponding to a normal distribution for the errors, with widths given by the corresponding error bars. 

Our choice of constraints consists of results for the entropy density \cite{Borsanyi:2013bia} and the second baryon susceptibility \cite{Borsanyi:2021sxv} from the Wuppertal-Budapest collaboration. These results were obtained at physical quark masses, with 2+1 flavors, and were extrapolated to the continuum.

A complication is that lattice QCD results at different temperatures can be correlated as a result of procedures such as the continuum extrapolation.\footnote{Correlations are visible in the way the lattice QCD error bars for neighboring points seem to line up.} 
We model these correlations with an extra parameter $-1<\Gamma <1$, quantifying correlations between neighboring points. 
Details on how we model the likelihood function can be found in Appendix~\ref{app:likelihood}.

\section{Results}

From the likelihood function and prior distributions, we can compute the posterior distribution $P(\vec\theta|\vec d)$ for any given set of parameters $\vec \theta$.  
We then use DE-MCMC to obtain posterior samples from this distribution. 
From these samples, we can obtain confidence intervals for model parameters and predictions. 

\begin{table*}
    \centering
    \begin{tabular}{|c||c|c||c|}
    \cline{1-3}
    PHA & \multicolumn{2}{|c||}{Posterior 95\% CI}&\multicolumn{1}{c}{}\\
    \cline{1-3}\hline
         Parameter & min & max & MAP  \\
    \hline\hline
        $\Lambda$ & 1089 MeV & 1190 MeV & 1129 MeV\\
    \hline
        $\kappa_5^2$ & 11.3 & 11.5 & 11.4\\
    \hline\hline
        $\gamma$ & 0.57 & 0.63 & 0.58\\
    \hline
        $b_2$ & 0.1 & 0.5 & 0.2\\
    \hline
        $b_4$ & -0.06 & -0.03 & -0.05\\
    \hline
        $b_6$ & 0.000 & 0.002 & 0.0007\\
    \hline\hline
        $c_1$ & -0.1 & 0.1 & 0.0\\
    \hline
        $c_2$ & 0.1 & 0.3 & 0.2\\
    \hline
        $c_3$ & 0.01 & 0.08 & 0.04\\
    \hline
        $d_1$ & 1.70 & 1.74 & 1.72\\
    \hline
      \phantom{(J)}    $d_2$ {(J)}  & 113 & 8068 & 1294\\
    \hline
    \end{tabular}\hspace{2cm}%
    \begin{tabular}{|c||c|c||c|}
    \cline{1-3}
    PA & \multicolumn{2}{|c||}{Posterior 95\% CI}&\multicolumn{1}{c}{}\\
    \cline{1-3}\hline
         Parameter & min & max & MAP \\
    \hline\hline
        $\Lambda$ & 862 MeV & 1043 MeV & 955 MeV\\
    \hline
        $\kappa_5^2$ & 11.3 & 11.5 & 11.4 \\
    \hline\hline
       $\gamma_1$ & 0.50 & 0.54 & 0.52 \\
       \hline
       $\gamma_2$ & 0.60 & 0.62 & 0.61 \\
    \hline
      $\Delta \phi_V$ & 1.6  & 2.1 & 1.8 \\
      \hline\hline
       $A$ & 0.369 & 0.374 & 0.371\\
    \hline
       $\phi_1$ & 0.000 & 0.025 & 0.002\\
    \hline
       \phantom{(J)}  $\delta\phi_1$  {(J)}  & 0.0001 & 0.0032 & 0.0003\\
    \hline
       $\phi_2$ & 2.1 & 2.3 & 2.2 \\
    \hline
       $\delta\phi_2$ & 0.65 & 0.73 & 0.69\\
    \hline
    \end{tabular}    
    \caption{Posterior 95\% confidence intervals (95\% CI) and maximum a posteriori (MAP) values for parameters of the PHA (left) and PA (right) models. The (J) marks parameters for which we have used \emph{Jeffreys} priors---i.e., prior distributions that are uniform over the logarithm of these parameters. MAP values are extracted by maximizing the likelihood. Corner plots showing two-dimensional marginal distributions of the posterior can be found in Appendix~\ref{app:posterior}. }
    \label{tab:posteriors}
\end{table*}

\subsection{Posterior distribution}

Table~\ref{tab:posteriors} shows the maximum a posteriori parameters and 95\% confidence intervals obtained in each \textit{Ans\"atze} for the holographic potentials.  
Both the PA and PHA models yield the same value and 95\% confidence interval for the gravitational constant $\kappa_5^2=11.4\pm 0.1$. 
The same is true for the correlation parameter  $\Gamma = 0.84^{+0.03}_{-0.06}$. 
Corner plots of the posterior distribution can be found in Appendix~\ref{app:posterior}.

Posterior samples for the zero-doping equation of state are shown as blue lines in Fig. \ref{fig:priors}, together with the lattice QCD results from Refs. \cite{Borsanyi:2013bia,Borsanyi:2021sxv} (red points). 
Even though, like the prior ones, these samples are shown {individually} as partially transparent lines, they concentrate in a clear-cut thin blue band, which roughly spans the entire region allowed by the lattice error bars.

\begin{figure}[h]
    \centering
    \includegraphics[width=\linewidth]{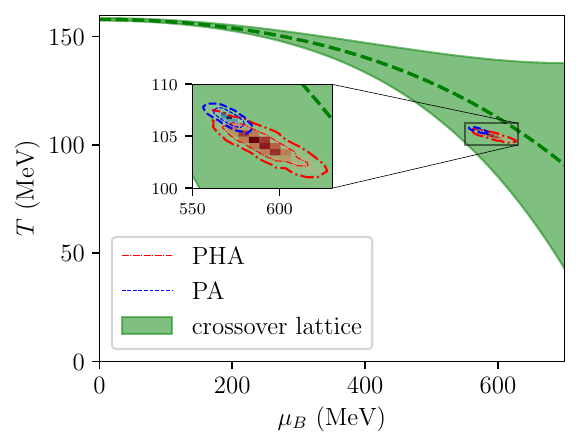}%
    \caption{Predictions for the location of the critical point on the $(T,~\mu_B)$-plane, based on the posterior distributions for the PHA model (red area) and the PA model (blue area). Also shown is the extrapolation of the lattice QCD transition line from Ref. \cite{Borsanyi:2020fev} (green band), based on the peak of the chiral susceptibility. 
Lines around confidence regions for the critical point location represent $68\%$ and $95\%$ confidence levels.
    }
    \label{Fig:CP2}
\end{figure}

\subsection{Critical point location}

Figure \ref{Fig:CP2} shows the predicted distributions for the critical point location from our Bayesian analysis for the PHA (red) and PA (blue)  \textit{Ans\"atze}, together with the corresponding $68\%$ and $95\%$ confidence levels. 
Differently from what was found for the prior samples, each critical point predicted within the posterior samples is located within a narrow region in $T$ and $\mu_B$. Moreover, the regions for the PA and PHA \textit{Ans\"atze} agree with each other, with overlapping $68\%$ confidence regions. 
This indicates that it is the lattice QCD results {at zero baryon density} that {provide the main} influence {on} the location of the critical point {in the holographic model}, regardless of the functional forms of the model potentials. 

Also shown in Fig.~\ref{Fig:CP2} is the extrapolation of the lattice QCD crossover line from Ref. \cite{Borsanyi:2020fev}, based on the peak of the chiral susceptibility (green band). 
The exact parametrization for the crossover temperature $T_{\textrm{cross}}$ from  \cite{Borsanyi:2020fev} is a quartic polynomial in $\hat\mu_B\equiv\mu_B/T_{\textrm{cross}}$, 
such that the decreasing $T_{\textrm{cross}}(\hat\mu_B)$ leads to a $\mu_B^{\textrm{cross}}(\hat\mu_B) \equiv \hat\mu_B\times T_{\textrm{cross}}(\hat\mu_B)$ which decreases with $\hat\mu_B$ for  $\hat\mu_B\gtrsim 2$.
To extrapolate this parametrization to values of $\mu_B\gtrsim 200$ MeV, we use it to find an expansion in powers of $\mu_B/T_{\textrm{cross}}(0)$ instead, which we also truncate at fourth order. 
This results in a tiny shift in the value of the quartic coefficient $\kappa_4$, with respect to the one presented in \cite{Borsanyi:2020fev}. 
Varying the coefficients $\kappa_2$ and $\kappa_4$ within uncertainties leads to the band shown in Fig.~\ref{Fig:CP2}.

It is evident that both $95\%$ confidence levels for the critical point are contained within the lattice extrapolation band. 
Since this did not have to be the case, we take this as a strong indication of the predictive power of the lattice QCD results, which strongly constrain the posterior distributions of critical points {of the holographic model} to a meaningful region of the phase diagram.
To enable the comparison to prior critical points, critical points drawn from the posterior are also shown in the insets on the rightmost panels of Fig.~\ref{fig:priors}.

Marginalizing the distribution in Fig.\ \ref{Fig:CP2} yields the following $95\%$ confidence intervals for the critical temperature $T_c$ and chemical potential $\mu_c$ :
\begin{eqnarray}
(T_c,\,\mu_{Bc})_{PHA}&=&(104\pm 3,\,589^{+36}_{-26})~\mathrm{MeV},\\
(T_c,\,\mu_{Bc})_{PA}&=&(107\pm 1,\, 571\pm 11)~\mathrm{MeV}.
\end{eqnarray}
These results are compatible with those obtained in Ref.\ \cite{Cai:2022omk}. %

\begin{figure}[t]
    \includegraphics[width=\linewidth]{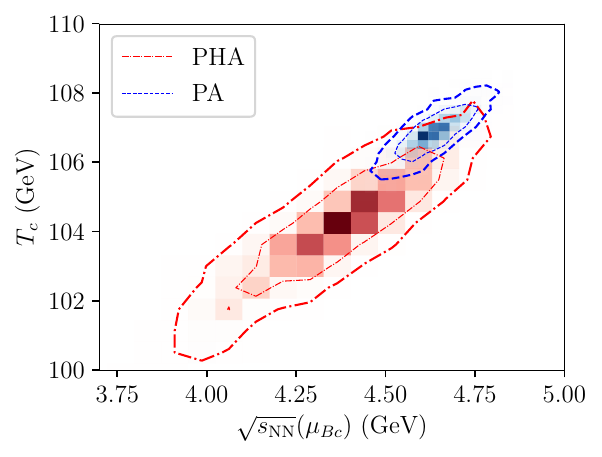}%
    \caption{%
Collision energy dependence of the baryon chemical potential for the critical point from the PA model (Blue, dashed lines) and the PHA model (red, dash-dotted lines). The predicted range for the center of
mass energy is  $\sqrt{s}=4.4\pm0.4$ GeV for the PHA model and $\sqrt{s}=4.6^{+0.2}_{-0.1}$ GeV for the PA one. 
The parametrization for $\mu_B(\sqrt{s})$ is taken from Ref.\ \cite{Vovchenko:2015idt}. 
Lines represent $68\%$ and $95\%$ confidence levels.
}
    \label{fig:sqrts}
\end{figure}

Next, we provide an estimate
for the center-of-mass energy, $\sqrt{s}$, in relativistic heavy-ion collisions that can potentially probe our holographic prediction for the location of the critical point. The analysis of mean hadron multiplicities within transport and statistical hadronization models provides the dependence of the variables $(T,~\mu_B)$ on $\sqrt{s}$ at the point where inelastic collisions cease to promote chemical equilibrium, i.e. the chemical freeze-out point \cite{Cleymans:2005xv,Andronic:2008gu,Vovchenko:2015idt,Alba:2020jir}. 
Additionally, the dependence on $\sqrt{s}$ can also be extracted from the
measurement of moments of net-particle distributions that can be directly compared with ratios of susceptibilities \cite{Alba:2014eba}. Taking into account the uncertainties in the location of the critical point from our analysis, we predict a range for
the center of mass energy of $\sqrt{s}=4.4\pm 0.4$ GeV for the PHA model,
and  $\sqrt{s}=4.6^{+0.2}_{-0.1}$ GeV for the PA model (see Fig. \ref{fig:sqrts}). 
 These results were obtained
by using the statistical hadronization model from Ref.\ \cite{Vovchenko:2015idt} for $\mu_B(\sqrt{s})$. 
Predictions for a holographic critical point have been previously presented in \cite{Critelli:2017oub,Li:2023mpv}. 

The prior and posterior samples used in our Bayesian analysis are publicly available in \cite{hippert_2024_13830379}, including predictions for the location of the critical point for each sample.  

\section{Conclusions}

In this manuscript, we presented the first prediction on the location of the critical point in the high-density and hot phase of strongly interacting quark-gluon matter, obtained through a Bayesian analysis constrained by first principle lattice QCD results at zero density. The analysis has been performed within the class of holographic EMD models. Different functional forms for the dilaton field potential and its coupling to the Maxwell field have been tested and constrained to reproduce the lattice QCD results for the entropy density and second-order baryon number susceptibility at $\mu_B=0$. While the prior distributions for all functional forms yield critical points that cover wide regions of the phase diagram, or no critical point at all, all posterior predictions for the critical point location collapse around $(T_c,\,\mu_{Bc})_{PHA}=(104\pm 3,\,589^{+36}_{-26})$ MeV and $(T_c,\,\mu_{Bc})_{PA}=(107\pm 1,\, 571\pm 11)$ MeV.  
 The two regions agree within 1 standard deviation, showing the ability of the lattice results {at zero baryon density} to strongly constrain the critical point location  {within the holographic model}. We predict that the collision energy needed to discover the critical point lies in the range: $\sqrt{s}=4.0-4.8$ GeV 
 , which is covered by the STAR Fixed Target program and could be explored at FAIR.\\

\section*{Acknowledgments}

We thank 
R.~Haas for discussions on computational aspects of this work. 
We also thank D.~Phillips for his insight on how to treat unknown error correlations and theory error, N.~Yunes for advice on Bayesian analyses, and  D.~Mroczek and G.~Nijs for fruitful discussions. 
This material is based upon work supported by the
National Science Foundation under grants No. PHY-2208724, PHY-1748958 and No. PHY-2116686 and in part by the U.S. Department of Energy, Office of Science, Office of Nuclear Physics, under Award Number DE-SC0022023, DE-SC0023861 and by the National Aeronautics and Space Agency (NASA) under Award Number 80NSSC24K0767. This work was supported in part by the National Science Foundation (NSF) within
the framework of the MUSES collaboration, under grant 
No. OAC-2103680. 
The authors also acknowledge support from the Illinois Campus Cluster, a computing resource that is operated by the Illinois Campus Cluster Program (ICCP) in conjunction with the National Center for Supercomputing Applications (NCSA), and which is supported by funds from the University of Illinois at Urbana-Champaign. 
R.R. acknowledges financial support by National Council for Scientific and Technological Development (CNPq) under grant number 407162/2023-2.

\bigskip

\renewcommand{\theequation}{A.\arabic{equation}} 
\setcounter{equation}{0}

\appendix

\section{LIKELIHOOD FUNCTION}
\label{app:likelihood}

The agreement between predictions $\vec p(\theta)$ of the model with parameters $\theta$ and lattice QCD results $\vec d$ is quantified by the likelihood function $\mathcal{L}(\vec\theta) \equiv P(\vec d| \vec \theta)$.

\begin{figure*}
    \centering
\includegraphics[width=\linewidth]{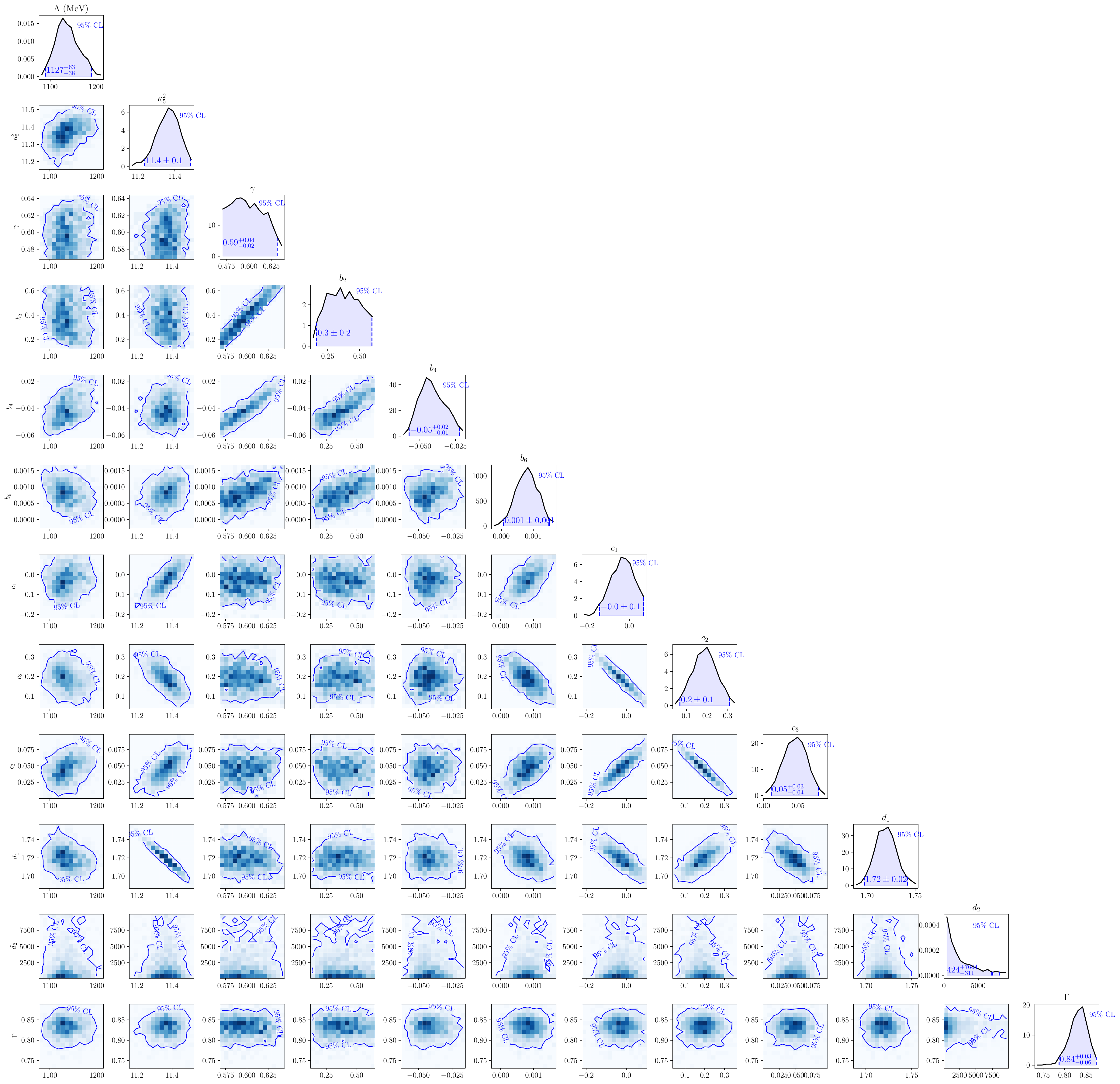}
    \caption{Marginalized \textit{a posteriori} probability distributions for pairs of parameters of the PHA model. Solid lines show 95\% confidence intervals. On the diagonal, marginalized one-parameter posterior distributions are also shown, along with the marginalized maximum a posteriori (MMAP) value and 95\% confidence interval for each parameter. MMAP values are extracted by maximizing the marginalized one-parameter posterior distributions.  }
    \label{fig:triangle-PHA}
\end{figure*}

\begin{figure*}
    \centering
    \includegraphics[width=\linewidth]{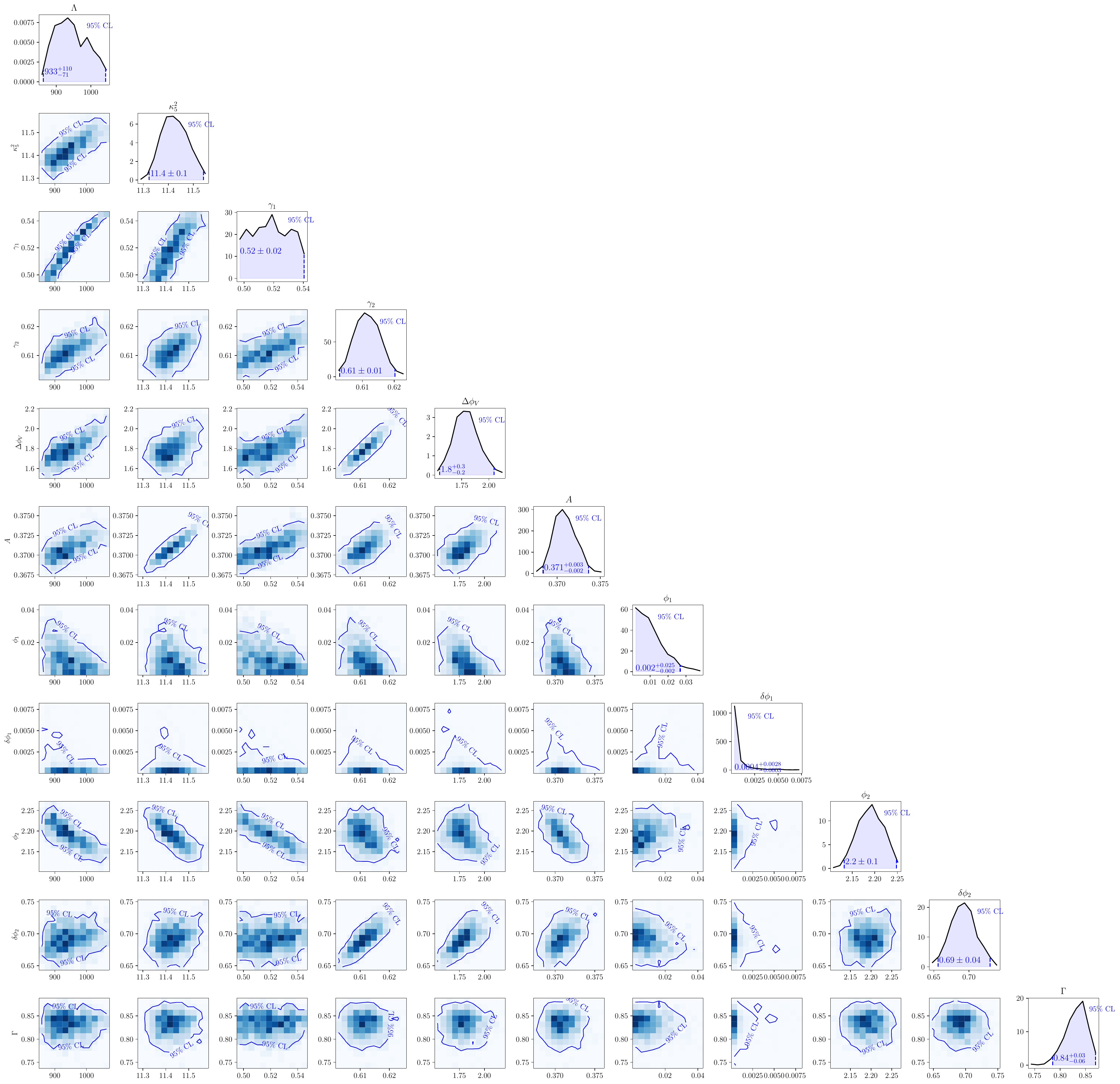}
    \caption{Marginalized a posteriori probability distributions for pairs of parameters of the PA model. Solid lines show 95\% confidence intervals. On the diagonal, marginalized one-parameter posterior distributions are also shown, along with the marginalized maximum a posteriori (MMAP) value and 95\% confidence interval for each parameter. MMAP values are extracted by maximizing the marginalized one-parameter posterior distributions. }
    \label{fig:triangle-PA}
\end{figure*}

We take a Gaussian likelihood 
\begin{widetext}
\begin{equation}
    \mathcal{L}(\vec \theta) = \frac{1}{\,\displaystyle\prod_{Q=s,\chi_2}\left[\left(\prod_i\sigma_i^{(Q)}\right)\sqrt{2\pi\det{\Lambda}}\right]}\exp\left\{-\frac{1}{2}\sum_{i,j}\sum_{Q=s,\chi_2}\frac{p_i^{(Q)}(\vec \theta)-d_i^{(Q)}}{\sigma_i^{(Q)}}\left[\Lambda^{-1}\right]_{ij}\frac{p_j^{(Q)}(\vec\theta)-d_j^{(Q)}}{\sigma_j^{(Q)}}\right\}, 
\end{equation}
\end{widetext}
where $\sigma_i^{(Q)}$, with $Q=s,\chi_2^B$ represent error bars for the different points from lattice QCD. 

The matrix $\Lambda$ is responsible for implementing correlations between neighboring points, by introducing an extra parameter $\Gamma \in (-1,1)$:
\begin{equation}
    \Lambda_{ij} = \Gamma^{|T_i-T_j|/\Delta T} ,
\end{equation}
where $\Delta T$ is the temperature step used to match all the points from lattice QCD. In principle, the determinant of $\Lambda$ should be computed only over the points where lattice results exist, since points are not always equally spaced and in the same interval for the two quantities in question. In practice, however, for simplicity, we assume there are $N=(T_{\textrm{max}}-T_{\textrm{min}})/\Delta T+1$ and take
\begin{equation}
    \det\Lambda\approx (1-\Gamma^2)^{N-1}.
\end{equation}

Finally, the logarithm of the likelihood becomes \cite{sivia1996data}
\begin{multline}
    \log\mathcal{L} = \frac{-1}{1-\Gamma^2}\left[ 
    (1+\Gamma^2)\,\zeta^2 - \Gamma\,\psi -\Gamma^2\,\phi
    \right]\\ - (N-1)\,\log(1-\Gamma^2) + \text{const.}\,,
\end{multline}
where
\begin{align}
    \zeta^2 &\equiv \frac{1}{2}\sum_{i}^N\sum_{Q=s,\chi_2}\left(\frac{p_i^{(Q)}(\vec \theta)-d_i^{(Q)}}{\sigma_i^{(Q)}}\right)^2,\\
    \psi &\equiv \sum_{i}^N\sum_{Q=s,\chi_2}\frac{p_i^{(Q)}(\vec \theta)-d_i^{(Q)}}{\sigma_i^{(Q)}}\frac{p_{i+1}^{(Q)}(\vec\theta)-d_{i+1}^{(Q)}}{\sigma_{i+1}^{(Q)}},\\
    \phi &\equiv 
     \frac{1}{2}\sum_{i=1,N}\sum_{Q=s,\chi_2}\left(\frac{p_i^{(Q)}(\vec \theta)-d_i^{(Q)}}{\sigma_i^{(Q)}}\right)^2    .
\end{align}

Remarkably, the posterior $95\%$ confidence interval obtained for the correlation strength  is of $\Gamma = 0.84^{+0.03}_{-0.06}$, for both the PHA and PA models.
This impressive agreement indicates that its value does not reflect the parametrization, but rather the lattice QCD error bars.

\begin{figure*}
    \centering
    \includegraphics[width=0.49\textwidth]{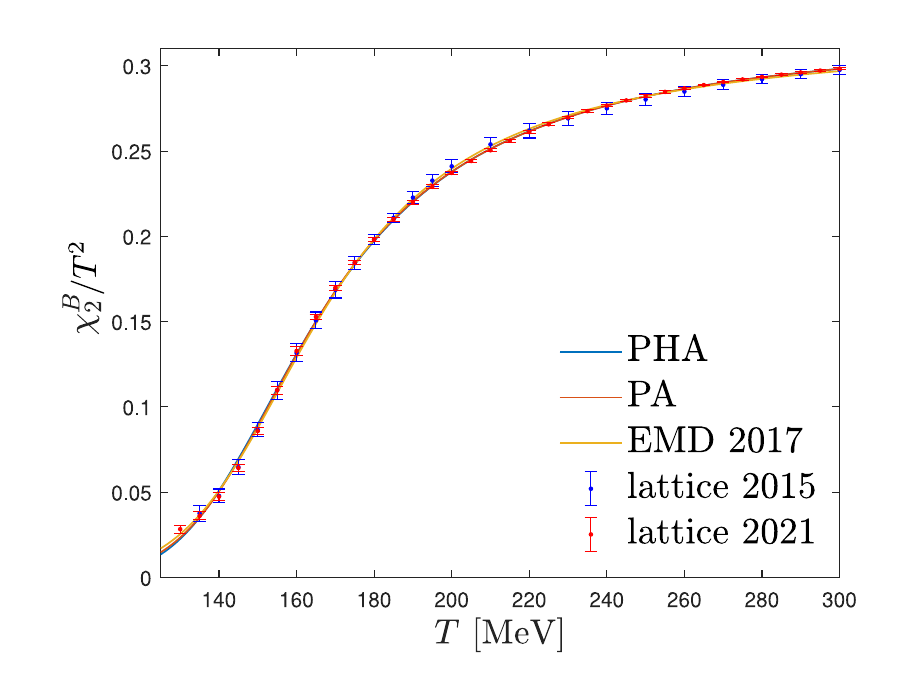}
    \includegraphics[width=0.49\textwidth]{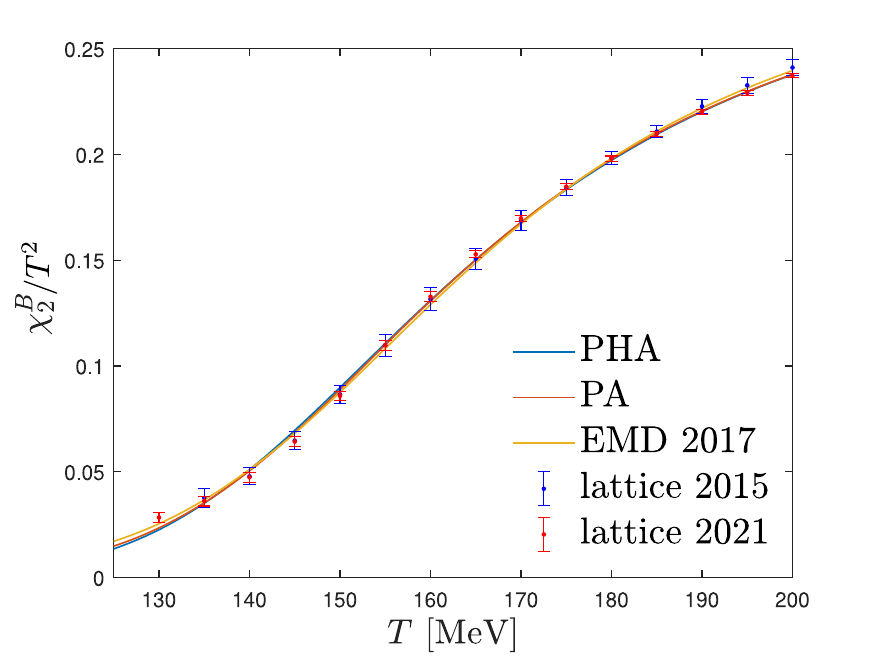}
    \caption{Second order baryon susceptibility $\chi_{2}^{B}$ 
    at zero chemical potential from the best fit to lattice data from Ref. \cite{Borsanyi:2021sxv} for PHA and PA. For comparison, we also show the previous lattice result for this susceptibility from Ref. \cite{Bellwied:2015lba} and the resulting fitting for our previous EMD model from Refs. \cite{Critelli:2017oub,Grefa:2021qvt}}
    \label{fig:chi2_comparison}
\end{figure*}

\begin{figure*}
    \centering
    \includegraphics[width=0.49\textwidth]{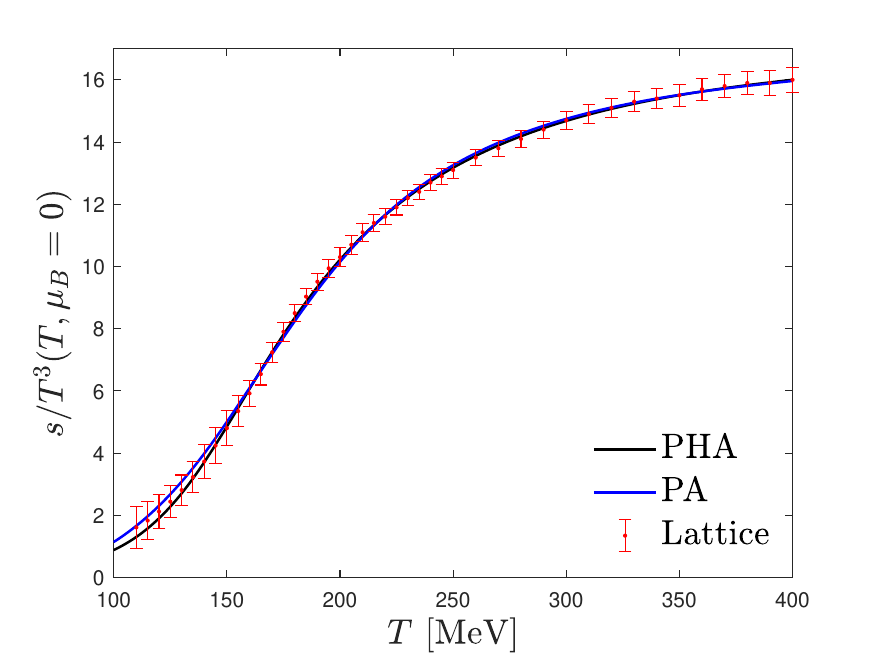}
    \includegraphics[width=0.49\textwidth]{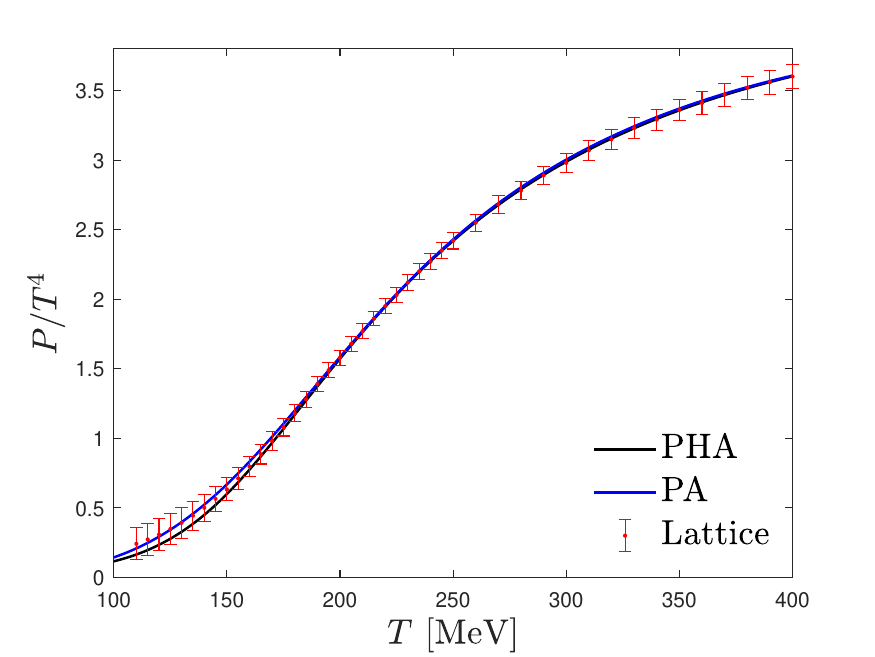}
    \includegraphics[width=0.49\textwidth]{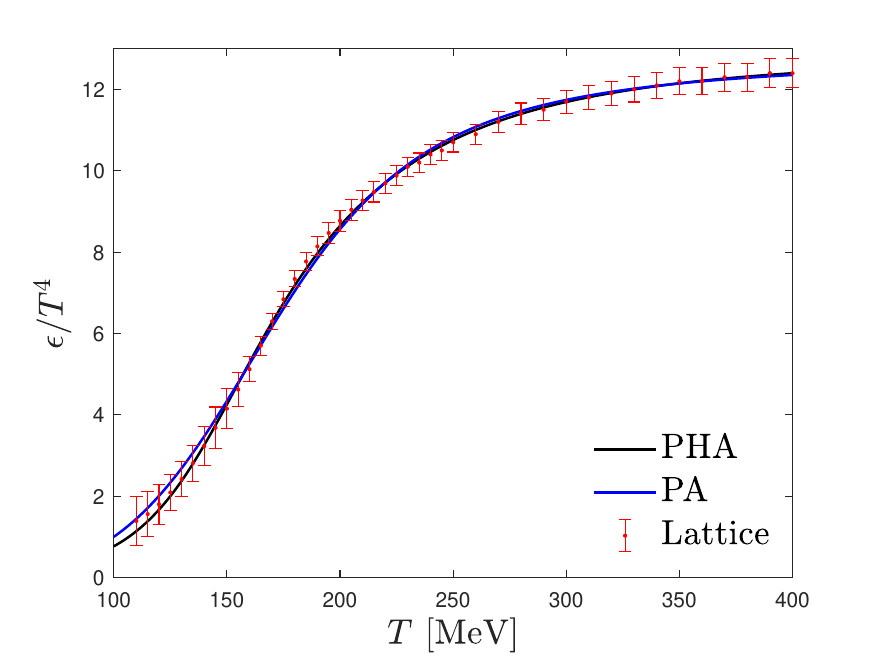}
    \includegraphics[width=0.49\textwidth]{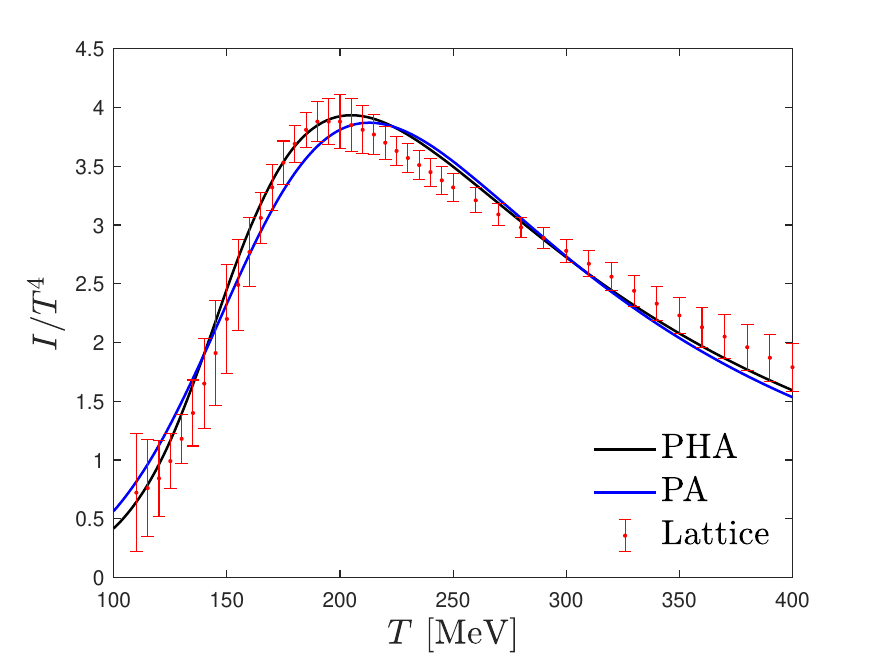}
    \caption{Comparison between the best fit for the PHA and PA models and the lattice equation of state at zero chemical potential from Ref. \cite{Borsanyi:2013bia}}
    \label{fig:EoS_mu0_comp}
\end{figure*}

\section{DE-MCMC ALGORITHM}
\label{app:demc}

To carry out our analyses, we had to overcome numerical challenges for the convergence of the MCMC. 
First of all, parameters in our posterior distributions turn out to be highly correlated. 
Furthermore, due to the small errors in lattice results, especially for the second-order baryon susceptibility, our posterior is strongly dominated by the likelihood, which delays the convergence of the MCMC. 

To overcome numerical challenges posed by correlations in a simple manner, we employ \emph{differential evolution} MCMC (DE-MCMC) \cite{speagle2019conceptual,2006S&C....16..239T}, with a simple Metropolis acceptance criterion. 
To avoid issues with local maxima of the posterior function, we employ the common strategy of increasing the relative step size every 10 iterations, so that each Monte Carlo chain can hop between local maxima \cite{2006S&C....16..239T}. 
To ensure all the chains reflect the same probability distribution so that DE-MCMC works in ideal conditions, enhancing the convergence of the algorithm, we also employ a simple sequential tempering of all the chains, choosing random starting samples from the previous step, according to the update in temperature, every time the MCMC temperature changes. 
After the  MCMC temperature reaches $1$, we let each chain evolve without interference.

\section{POSTERIOR DISTRIBUTIONS}
\label{app:posterior}

After convergence is found, we collect samples. To ensure statistical independence between samples, we ``thin out'' the data (i.e., skip samples) according to the measured correlation time. 

The resulting posterior distributions for the parameters of the PHA and PA models are illustrated in the corner plots in Figs.~\ref{fig:triangle-PHA} and \ref{fig:triangle-PA}, respectively, where 95\% confidence levels are also shown. 

The same 95\% confidence intervals are shown in Table~\ref{tab:posteriors}. However, in that table, maximum a posteriori parameters are extracted from the point of maximum likelihood, while in Figs.~\ref{fig:triangle-PHA} and \ref{fig:triangle-PA}, marginalized maximum a posteriori parameters are extracted from the marginalized distribution for each parameter.

\begin{figure*}
    \centering
    \includegraphics[width=0.4\textwidth]{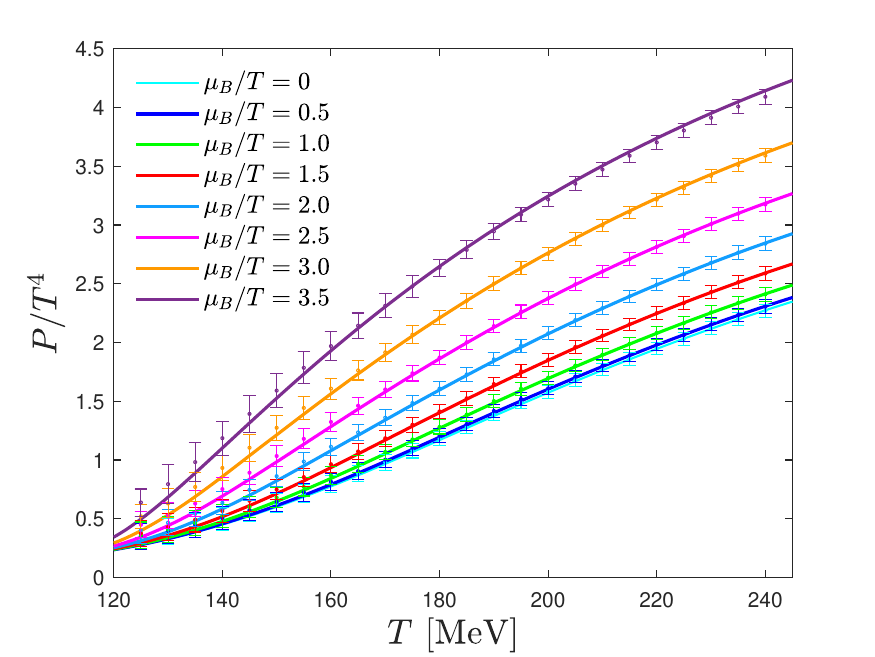}   \includegraphics[width=0.4\textwidth]{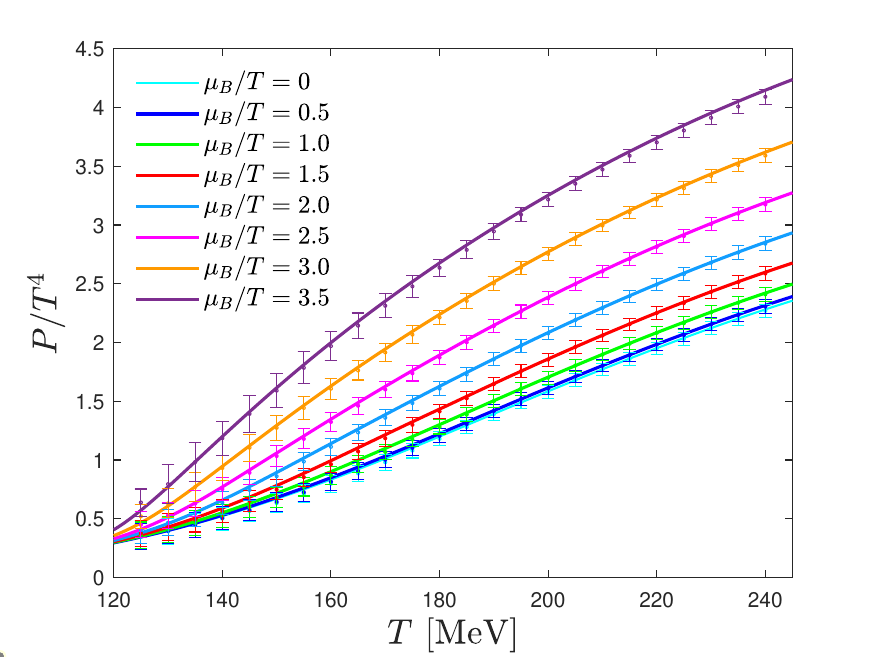}
    \newline
    \includegraphics[width=0.4\textwidth]{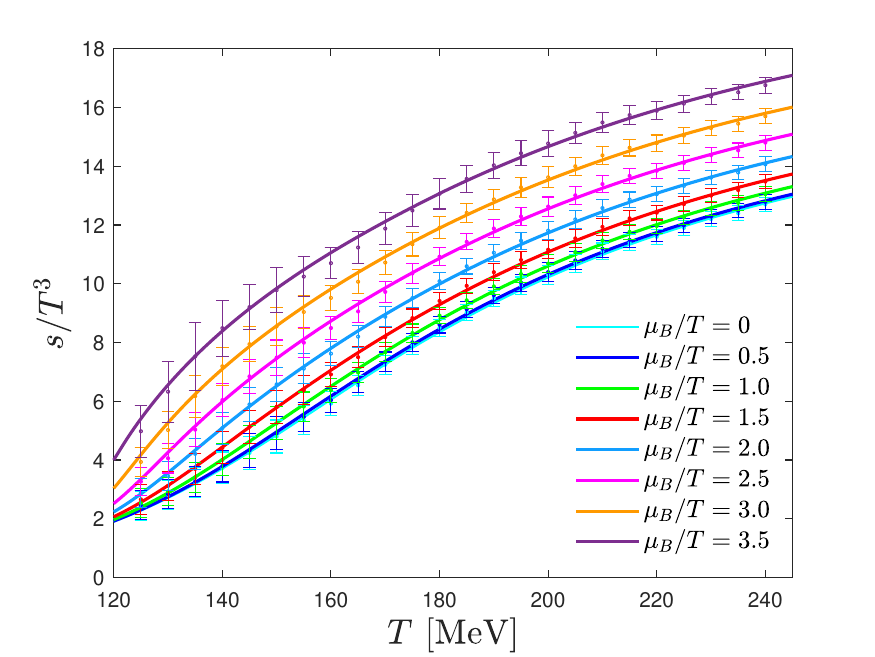}   \includegraphics[width=0.4\textwidth]{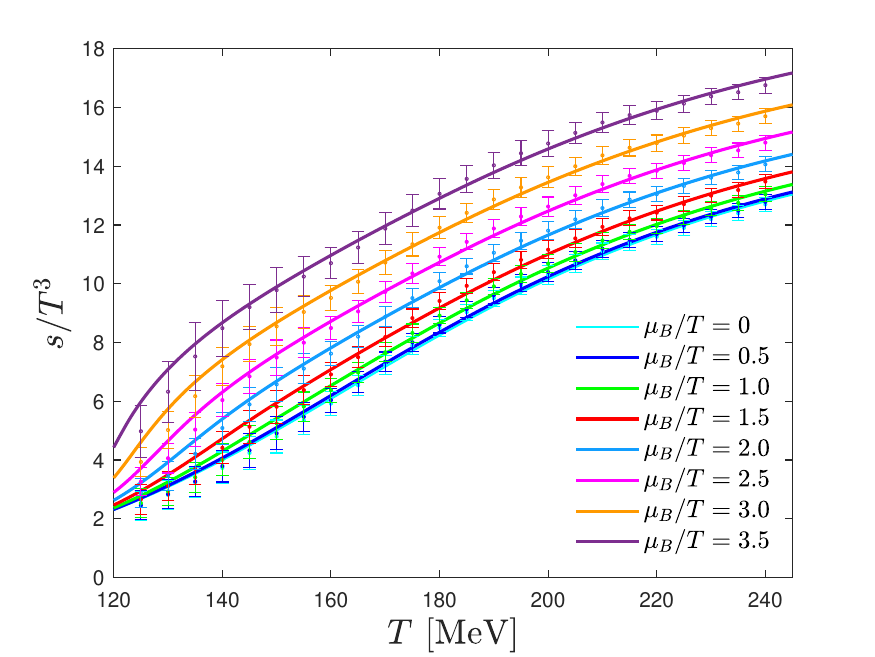}
    \newline
    \includegraphics[width=0.4\textwidth]{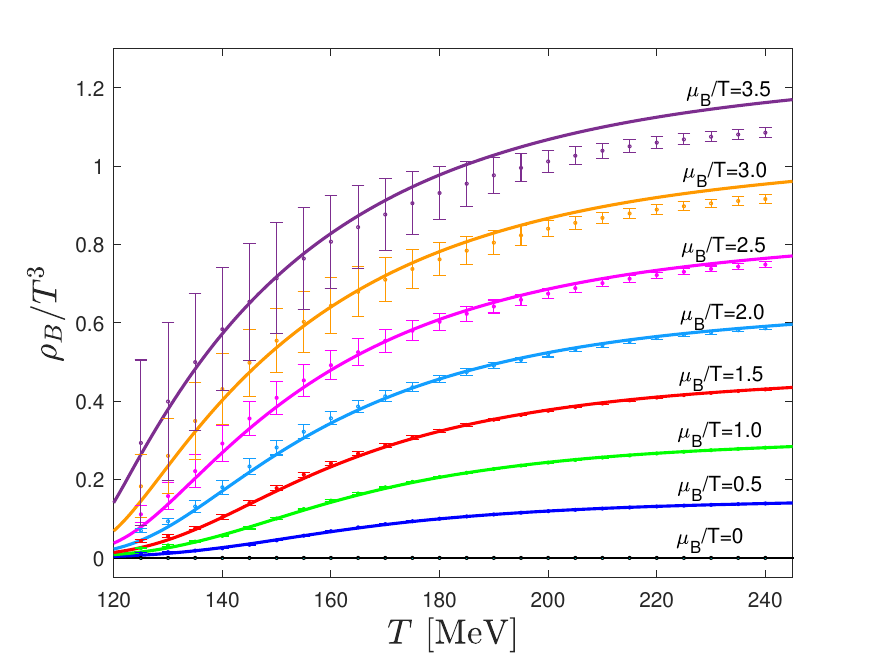}   \includegraphics[width=0.4\textwidth]{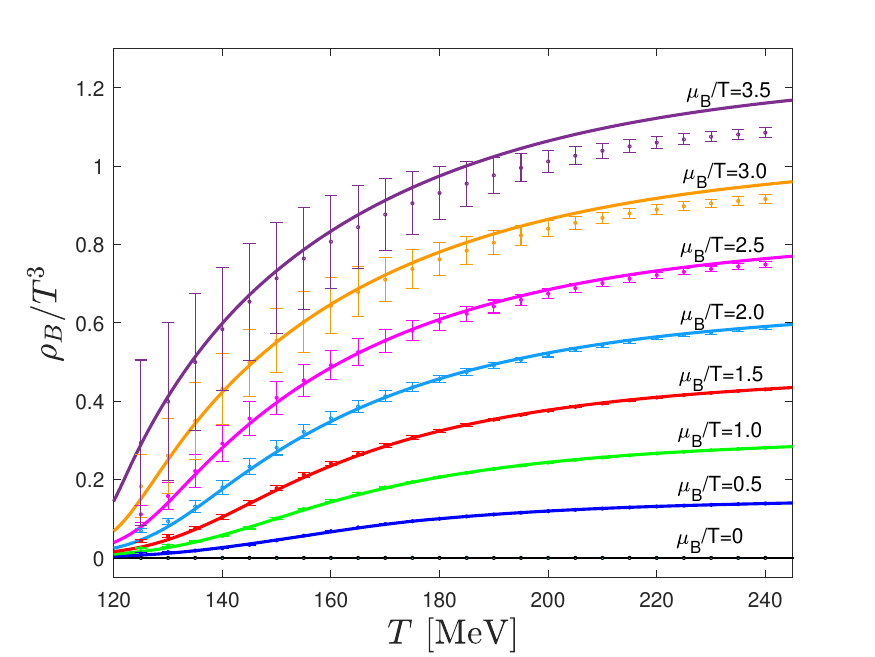}  
   \newline
    \includegraphics[width=0.4\textwidth]{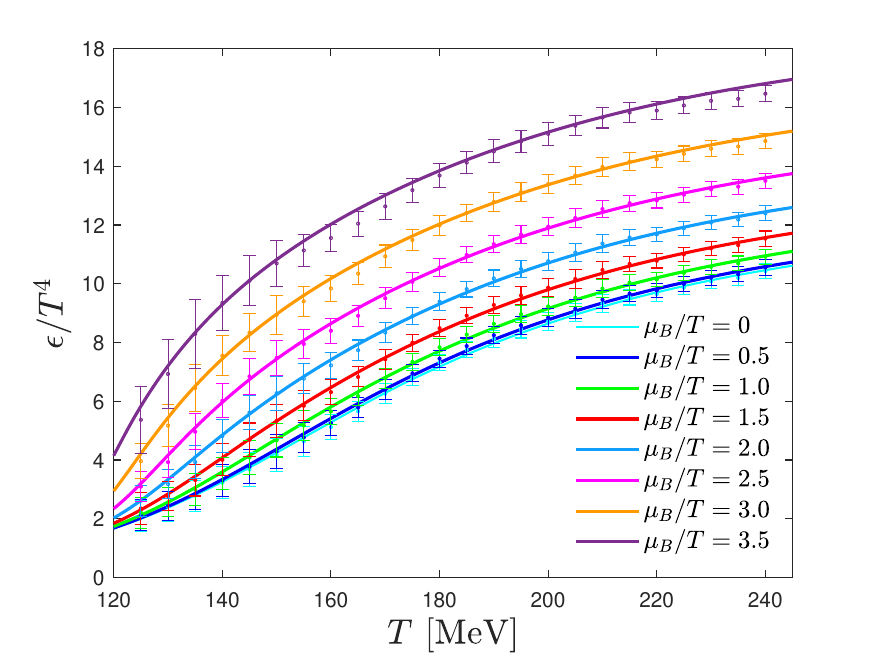} \includegraphics[width=0.4\textwidth]{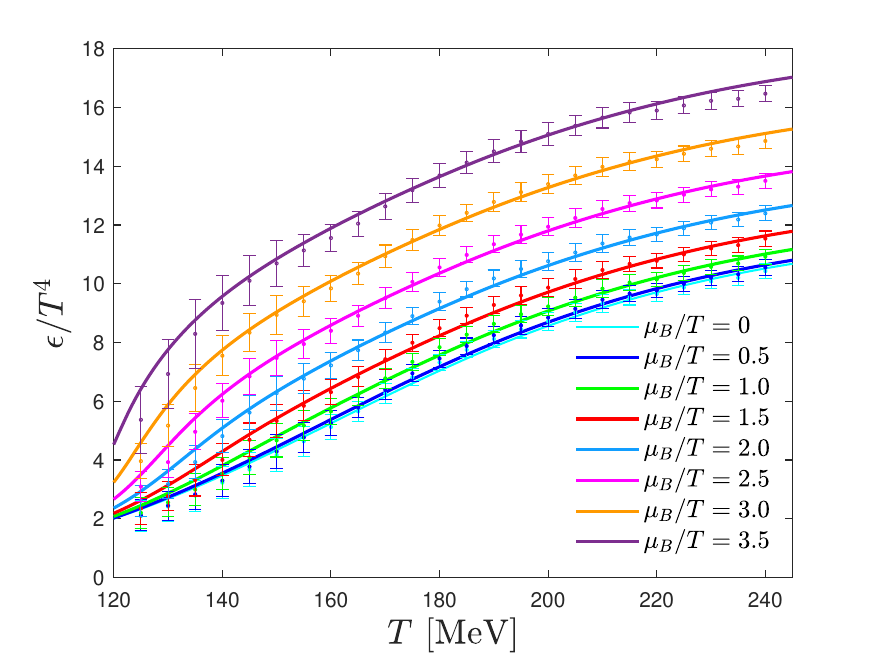} 
    \newline    
    \caption{Comparison between the Equation of State at finite density obtained from the best fit to lattice data at zero chemical potential for the PHA model (left) and the PA one (right), and the state-of-the-art lattice QCD results from \cite{Borsanyi:2021sxv}.}
    \label{fig:finite_density_EoS}
\end{figure*}

\section{COMPARISON WITH LATTICE QCD RESULTS}
\label{app:lattcomp}
 In this section, we compare state-of-the-art lattice QCD results to the corresponding holographic postdictions for the PA and PHA models, with the best-fit parameters from the Bayesian analysis. In Fig. \ref{fig:chi2_comparison} we present the comparison of the second-order baryon susceptibility at $\mu_{B}=0$ between the holographic model and the Lattice QCD results from Ref. \cite{Borsanyi:2021sxv}, which were used to constrain the PA and PHA models. For further contrast, we also show the previous lattice data from Ref. \cite{Borsanyi:2021sxv} and the previous holographic result from Refs. \cite{Critelli:2017oub,Grefa:2021qvt}. It is worth noticing that the error bars for the lattice susceptibility have reduced, and the holographic susceptibility from both models almost overlaps completely.

In Fig. \ref{fig:EoS_mu0_comp}, we also show the holographic  results in the PA and PHA models for the equation of state at zero chemical potential, compared to the lattice results from  \cite{Borsanyi:2013bia}. It is important to remark that the entropy density, together with the second-order baryon susceptibility, is the quantity that is matched to the lattice result at zero chemical potential, whereas other thermodynamic quantities are computed directly from thermodynamic identities. 

One can observe minor differences between the best PHA and PA models. The PA model gets closer to the lattice data at very low T, but then the PHA model gets a better agreement after the third lowest temperature point, although both are within error bars. At some point, both results completely overlap in the case of the entropy density and pressure. However, for the trace anomaly, the PHA model is in better agreement with the corresponding lattice results. This is not too surprising, since the lattice error bars are large at low $T$, which means that those data points are the least constrained in the Bayesian analysis. Thus, the models with the highest likelihood are the ones that fit the high $T$ error bars, which are much smaller. At high temperatures, the results from both PA and PHA completely overlap.

However, the most important comparison is the one between the predicted holographic equation of state and the most recent lattice results for the QCD equation of state at finite temperature and chemical potential. This comparison with the lattice results from Ref. \cite{Borsanyi:2021sxv} is shown in Fig. \ref{fig:finite_density_EoS}. One can note that the holographic prediction for the entropy density in the PHA (left) and PA (right) models are in numerical agreement with the lattice points, although, in the case of the pressure and energy density, the holographic results start to deviate from the lattice prediction when $\mu_{B}/T\ge 3.5$. Similarly to what was reported in our previous work \cite{Grefa:2021qvt}, the holographic baryon density is in agreement with the lattice results up to $T\approx190$ MeV for all values of $\mu_{B}/T$.
Additionally, one difference we observe 
is the location of the critical point from the best fit (i.e, maximum likelihood parameters) for each model. The predicted critical point for the PHA model with the highest likelihood is found at $T_c=103.45$ MeV, and $\mu_{Bc}=599$ MeV, whereas for the PA case, the critical point is located at $T_c=106.72$ MeV, $\mu_{Bc}=573$ MeV.

\bibliographystyle{unsrturl}
\bibliography{reference.bib,noninspire.bib}

\end{document}